\documentclass[twocolumn]{article}
\usepackage[margin=0.7in]{geometry}
\usepackage{graphicx}
\usepackage{amsfonts}
\usepackage{amssymb}
\usepackage{array}
\usepackage{amsmath}
\usepackage{color}
\usepackage{verbatim}
\usepackage{slashed}
\usepackage{subfigure}
\usepackage{xcolor}
\usepackage{bbold}
\usepackage{mathtools}
\usepackage{bbm}
\usepackage{amsthm}
\usepackage{authblk}
\usepackage{cite}

\makeatletter
\renewcommand{\maketag@@@}[1]{\hbox{\m@th\normalsize\normalfont#1}}%
\makeatother

\theoremstyle{plain}
\newtheorem*{thm}{Theorem}
\theoremstyle{definition}

\title{Entropic lower bound for distinguishability of quantum states}
\begin{document}
\author[1]{Seungho Yang}
\author[1,2]{Jinhyoung Lee}
\author[1]{Hyunseok Jeong}
\affil[1]{Center for Macroscopic Quantum Control, Department of Physics and Astronomy, Seoul National University, Seoul, 151-742, Korea}
\affil[2]{Department of Physics, Hanyang University, Seoul 133-791, Korea}

\maketitle
\abstract{
For a system randomly prepared in a number of quantum states, we present a lower bound for the distinguishability of the quantum states, that is, the success probability of determining the states in the form of entropy. When the states are all pure, acquiring the entropic lower bound requires only the density operator and the number of the possible states. This entropic bound shows a relation between the von Neumann entropy and the distinguishability.
}

\section{introduction}
Quantum mechanics does not allow to determine the state of a system by measuring a single copy of the ensemble. Nevertheless, if some prior information is known, it is possible to guess the state with a certain degree of confidence even by a single measurement. Given some prior information, to what extent quantum states can be distinguished is an intriguing issue from both fundamental and practical points of view.
For example, this problem is closely related to efficiencies of quantum communication~\cite{Helstrom76,Holevo00,Scarnai09}.
It is known that the imperfect distinguishability plays a crucial role in the security of quantum cryptography~\cite{Scarnai09}.

There are different approaches to the distinguishability of quantum states~\cite{Barnnet09,Audenaert07,Bergou05,Dusek02}. In the minimum-error discrimination problem~\cite{Barnnet09}, a set of known quantum states and preparation probabilities are given, and one aims to distinguish the states with the optimal probability of success. The optimal success probability is an operationally well-defined measure for the distinguishability of given states. However, it is a highly demanding quest to find its analytical solution for general sets of states, and the solution is only known for the sets of two states~\cite{Helstrom76}. Instead, upper bounds~\cite{Hayashi,Montanaro08,Qiu08-1,Qiu08-2,Tyson09} and lower bounds \cite{Kholevo79,Montanaro07, Barnum02} have been provided to estimate the optimal probability. On the other hand, there have been studies of distinguishability between unknown quantum states using programmable machines~\cite{Bergou05,Dusek02}. In this case, ancillary systems prepared in each of the unknown states are provided as an input of the machine.

One may pay attention to the von Neumann entropy as a quantity related to the distinguishability in light of the capacity of a quantum state for embodying quantum information.
When a system is probabilistically prepared in one of a certain number of quantum states, its state of the statistical mixture is described by a density operator. According to the quantum source theorem~\cite{Schumacher95}, the von Neumann entropy of the system, which is given as a function of the density operator, represents the capacity of the mixed state to (asymptotically) carry quantum information. As discussed in Ref.~\cite{Jozsa00}, one may relate this capacity to the concept of distinguishability because more information could be carried when each state is more distinguishable. For this reason, Jozsa and Schlienz considered the von Neumann entropy as a measure for distinguishability~\cite{Jozsa00}.
However, it is not known whether this kind of distinguishability is linked to the actual ability to distinguish quantum states by measurements, namely the success probability (distinguishability will henceforth refer to the success proabability). It seems that, at least, the von Neumann entropy cannot pinpoint the success probability of distinguishing given quantum states. This is because the von Neumann entropy is determined only  by the density operator, and the density operator of a system can take arbitrarily many decompositions in general; thus it does not contain information on which states the system could have been prepared in.

We here present a lower bound for the distinguishability, i.e., the optimal success probabilities of distinguishing between quantum states, as a function of entropies of the system. For a system prepared in one of $N$ pure states, we also present a reduced form of the entropic bound which requires the density operator and the number of possible states $N$ for its evaluation.
It reveals a relation between the von Neumann entropy and the distinguishability; the larger von Neumann entropy guarantees the better distinguishability.

\section{General formulation}
 Consider a quantum system prepared in one of $N$ quantum states with some probabilities. We denote them by $\{p_{x},\rho_{x}\}_{x=1}^{N}$. One wishes to identify which state the system has been prepared in or, equivalently, to identify the value of $x$. The value of $x$ is determined, using a generalized measurement described by the measurement operator $\{\hat{M}_{x}\}_{x=1}^{N}$. Therefore, the probability of correctly identifying $x$ for a given $x$ is $\text{Tr}[\hat{M}_{x}\rho_{x}]$, and the expected success probability is
 \begin{equation}
  P_{\text{s}}=\sum_{x=1}^{N} p_{x}\text{Tr}\left(\hat{M}_{x}\rho_{x}\right).
 \end{equation}
 When maximized over all measurements, it becomes the optimal success probability, which is denoted by $P_{\text{s}}^{*}$. It is the quantity which we consider as the degree of the distinguishability of quantum states. As already mentioned, however, the analytical form of $P_{\text{s}}^{*}$ is known only for the two-state case~\cite{Helstrom76}.

 An equivalent way of describing the scenario is to consider a classical-quantum system $XQ$ in the state,
\begin{equation}\label{c-q state}
\sigma_{XQ}=\sum_{x}p_{x}{\vert}x\rangle\langle x{\vert}\otimes \rho_{x},
\end{equation}
 where the indices $x$'s are encoded. Namely, one is given the quantum system $Q$ and wishes to determine the value of $x$ by measuring $Q$.
 In terms of entropic quantities, $X$ has uncertainty quantified by Shannon entropy $H(X)=-\sum p_{x}\log p_{x}$, and it has a correlation with the quantum system.

 One may expect an entropic lower bound from the intuition that the correlation of $Q$ with $X$ would enhance the distinguishability of the quantum states. For better understanding, let us first consider a fully classical case~\cite{Nayak06}, where the quantum system $Q$ is replaced with a classical system $Y$ and $\rho_{x}=\sum_{y}p(y\vert x)\vert y\rangle\langle y\vert$.
 Assume that we are given $Y$ and wish to determine the value of $x$ from $y$.
  For a given $y$, the most probable $x$ is the one that gives the maximum conditional probability, $\max_{x} p(x\vert y)$. Therefore, the optimal success probability is attained by choosing them for all $y$, and it is given as $P_{\text{s}}^{*}=\sum_{y}p(y)\max_{x} p(x\vert y)$. It can be lower bounded in terms of the correlation between $X$ and $Y$ as follows.
 \begin{equation*}
 \begin{split}
 P_{\text{s}}^{*}&=\sum_{y}p(y)\max_{x} p(x\vert y)
 \geqslant\sum_{y}p(y)\sum_{x}p(x\vert y)^{2}\\
 &=\sum_{x,y}p(x,y)p(x\vert y)
 =\sum_{x,y}p(x,y)2^{\log p(x\vert y)} \\
 &\geqslant 2^{\sum_{x,y}p(x,y)\log p(x\vert y)}.
 \end{split}
 \end{equation*}
  The first inequality follows by taking the average of $p(x\vert y)$ over $p(x\vert y)$, and the second one follows from the concavity of the exponential function. The exponent in the last line is equal to the conditional Shannon entorpy $H(X\vert Y)$ so that
 \begin{equation}\label{c_bound}
P_{\text{s}}^{*}\geqslant 2^{-H(X\vert Y)}=2^{-H(X)+H(X:Y)},
\end{equation}
 where $H(X:Y)$ is the classical mutual information between $X$ and $Y$. Therefore, with assistance from the random variable $Y$ having the amount of correlation $H(X:Y)$, one can guess the random variable $X$ with probability (in logarithm) at least $-H(X)+H(X:Y)$.

\section{Entropic lower bound for the distinguishability of quantum states}
  For the quantum case, we can still obtain a random variable $\tilde{Y}$ by applying a measurement (i.e., the outcome of the measurement can be considered as a random variable), and applying \eqref{c_bound} gives $\log P_{\text{s}}^{*}\geqslant -H(X\vert\tilde{Y})$. However, this bound can be further sharpened by using the quantum entropies. For a quantum system prepared in a density operator $\chi_{A}$, the von Neumann entropy $S(A)$ is defined as $S(A)=S(\chi_{A})=-\text{Tr}[\chi_{A}\log\chi_{A}]$. The conditional von Neumann entropy of a bipartite system $AB$ is defined as $S(A\vert B)=S(AB)-S(B)$. Similarly, $I(A:B)=S(A)+S(B)-S(AB)$ defines the von Neumann mutual information~\cite{Adami97}. In terms of the quantum entropies, we present the quantum entropic lower bound.

 \begin{thm} For a set of quantum states with preparation probabilities $\{p_{x},\rho_{x}\}_{x=1}^{N}$, the optimal success probability of distinguishing the quantum states, $P_{s}^{*}$, is lower bounded as
 \begin{equation}\label{q_bound1}
 \begin{split}
 P_{\text{s}}^{*}\geqslant 2^{-S(X\vert Q)}&=2^{-H(X)+I(X:Q)} \\&=2^{-H(\vec{p})+S\left(\sum p_{x}\rho_{x}\right)-\sum p_{x}S(\rho_{x})}.
 \end{split}
 \end{equation}
 \end{thm}

  \emph{Proof.~}For the proof, we employ the conditional min-entropy~\cite{Renner04}, which has many applications in quantum cryptography. The conditional min-entropy $S_{\text{min}}(A\vert B)$ of a system $AB$ in a state $\rho_{AB}$ is defined as
   \begin{small}
   \begin{equation*}
   S_{\text{min}}(A\vert B)=-\inf_{\sigma_{B}\in \pi_{B}} \inf \left\{\lambda \in\mathbb{R}: \rho_{AB}\leqslant 2^{\lambda} (\mathbb{1}_{A}\otimes \sigma_{B})\right\}
   \end{equation*}
   \end{small}
   where $\pi_{B}$ denotes the set of all quantum states of the subsystem $B$.
   We derive the lower bound~\eqref{q_bound1} from two properties of the conditional min-entropy. First, the conditional min-entropy has an operational meaning that for the classical-quantum states in \eqref{c-q state}, the logarithm of $P_{\text{s}}^{*}$ is equal to the negative conditional min-entropy (see Ref.~\cite{Koenig09} for the details). Therefore, for the classical-quantum state in \eqref{c-q state},
  \begin{equation}\label{op. equality}
  \log P^{*}_{\text{s}}=-S_{\text{min}}(X\vert Q).
  \end{equation}
It has also been shown in Ref.~\cite{Tomanichel09} that the conditional min-entropy is always less than or equal to the conditional von Neumann entropy, so we have
 \begin{equation}\label{monotonicity}
 S_{\text{min}}(A\vert B) \leqslant S(A\vert B).
 \end{equation}
Using Eqs.~\eqref{op. equality} and ~\eqref{monotonicity}, we obtain
 \begin{equation}
 \log P_{\text{s}}^{*} \geqslant -S(X\vert Q),
 \end{equation}
  which is equivalent to the first line in \eqref{q_bound1}. On the other hand, the conditional von Neumann entropy satisfies the chain rule, $S(X\vert Q)=S(X)-S(XQ)$. It enables the lower bound to be written as a function of entropies of the system $Q$.  The von Neumann entropies of $XQ$ and $X$ are evaluated as $S(XQ)=S(\sum p_{x}\rho_{x})-\sum p_{x}S(\rho_{x})$ and $S(X)=H(\vec{p})$. It then follows that $S(X\vert Q)=H(\vec{p})-S(\sum p_{x}\rho_{x})+\sum p_{x}S(\rho_{x})$, and this gives the second line in \eqref{q_bound1}.~\hfill $\square$


   Let us take a closer look into the form of the lower bound. Its form is exactly of \eqref{c_bound}, but only $H(X:Y)$ is replaced with $I(X:Q)$. The quantum mutual information is known to capture all the correlations including both classical and quantum parts~\cite{Henderson01,Groisman05}, so it is considered as a measure of the total correlations. Hence, we see that the lower bound is increased by the total correlation.
   For the classical-quantum state $\sigma_{XQ}$, we obtain $I(X:Q)=S(\sum p_{x}\rho_{x})-\sum p_{x}S(\rho_{x})$, and Holevo's theorem~\cite{Holevo73,Nielsen00} implies that $H(X:\tilde{Y})\leqslant S(\sum p_{x}\rho_{x})-\sum p_{x}S(\rho_{x})=I(X:Q)$ for any measurement on the system $Q$ ($\tilde{Y}$ is the outcome of a measurement on $Q$). Their minimum difference between $I(X:Q)$ and $H(X:\tilde{Y})$, i.e., $I(X:Q)-\max H(X:\tilde{Y})$, is equal to quantum discord \cite{Ollivier01} of $\sigma_{XQ}$.

   On the assumption that the system is prepared in one of $N$ pure states $\{p_{x},\psi_{x}\}_{x=1}^{N}$, the entropic bound can be reduced to a form that only requires the density operator of the system and the number of the possible states for its evaluation. Using $H(\vec{p})\leqslant \log N$ and $S(\psi_{x})=0$, we have
  \begin{equation}\label{q_bound2}
    \begin{split}
    P_{\text{s}}^{*}\geqslant \frac{2^{S\left(\sum p_{x}\vert \psi_{x}\rangle\langle\psi_{x}\vert\right)}}{N}.
    \end{split}
  \end{equation}

 Therefore, we see that the larger von Neumann entropy entropy guarantees the better distinguishability. Notwithstanding the missing information on the component states and the preparation probabilities, the density operator of a system alone can provide a lower bound for distinguishability of $N$ pure states. Note that when many copies of quantum systems are prepared in the same way, the density operator can be obtained using the state tomography, but it is impossible to guess the component states. The state-discrimination machine with unknown quantum states as an input~\cite{Bergou05,Dusek02} is the case where it is required to distinguish between unknown quantum states. 

 \section{Other lower bounds and examples}

 In this section, we compare the entropic lower bound to other previously known bounds. One is the lower bound given by the square-root measurement~\cite{Kholevo79}, and it is known to be optimal for many cases in which the solutions are known~\cite{Barnnet09}. The measurement operators of the square-root measurement $\{\pi_{i}\}_{i=1}^{N}$ are given as
 \begin{equation*}
\hat{\pi}_{x}= p_{x}\rho^{-\frac{1}{2}}\rho_{x}\rho^{-\frac{1}{2}},
\end{equation*}
where $\rho=\sum_{x}p_{x}\rho_{x}$. Therefore, a lower bound by the square-root measurement is given as
\begin{equation}\label{srm}
P_{\text{s}}^{*}\geqslant \sum_{x=1}^{N} p_{x}\text{Tr}\left(\rho_{x} \hat{\pi}_{x}\right).
\end{equation}

Another one is the pairwise-overlap bound. For an ensemble of pure states $\{p_{x},\psi_{x}\}_{x=1}^{N}$, it has been given in Ref.~\cite{Montanaro07} as
 \begin{equation} \label{pairwise bound}
 P_{\text{s}}^{*}\geqslant \sum_{i=1}^{N}\frac{p_{i}^{2}}{\sum_{j=1}^{N}p_{j}\vert\langle\psi_{i}\vert\psi_{j}\rangle\vert^{2}}.
 \end{equation}
This was derived as a lower bound for the square-root measurement bound, so it is always less than or equal to the bound by the square-root measurement. It provides an analytic form of lower bounds in terms of the pairwise overlaps.

\begin{figure*}
\centering
\subfigure{\includegraphics[width=0.45\textwidth]{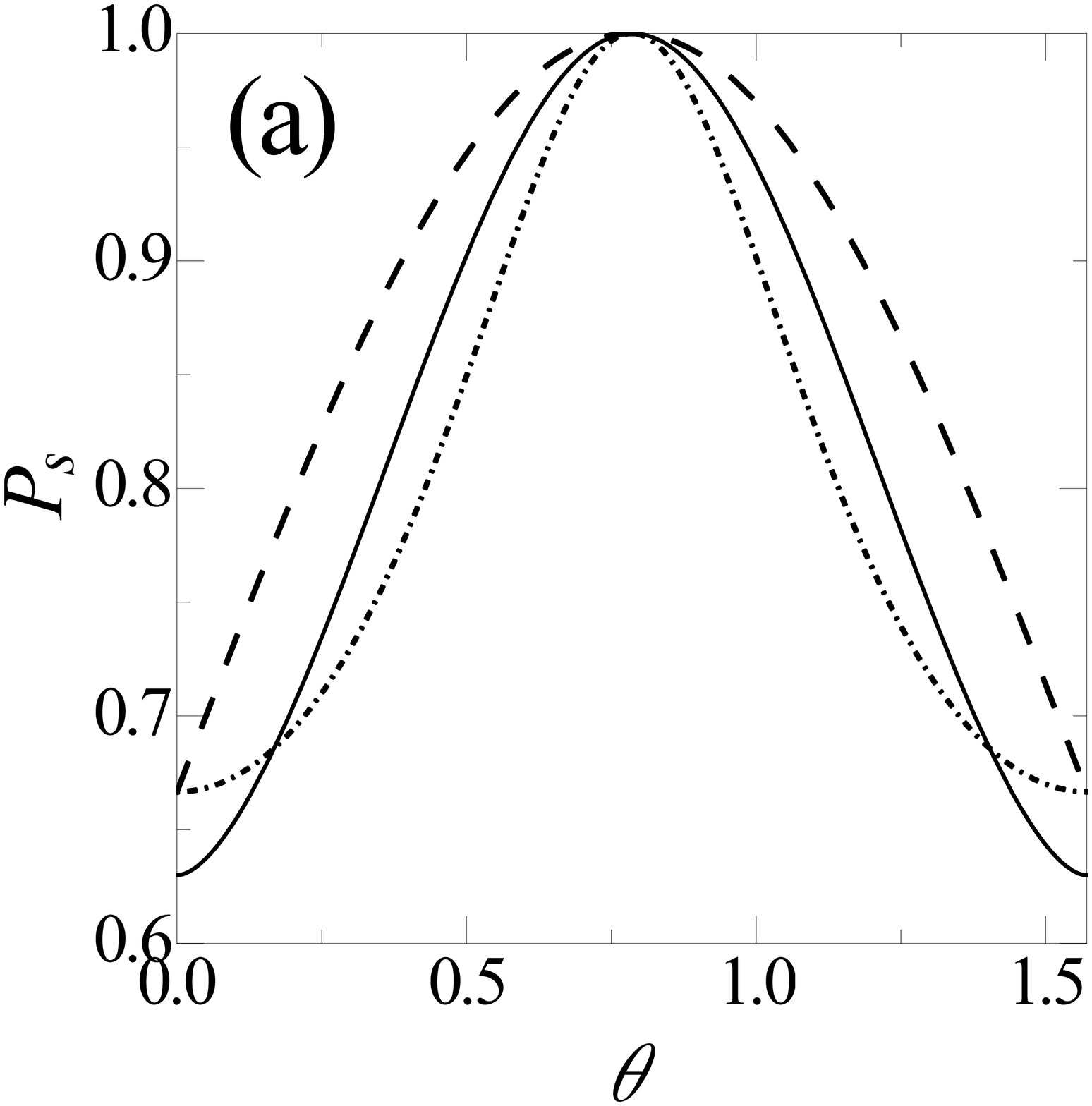}}
\subfigure{\includegraphics[width=0.45\textwidth]{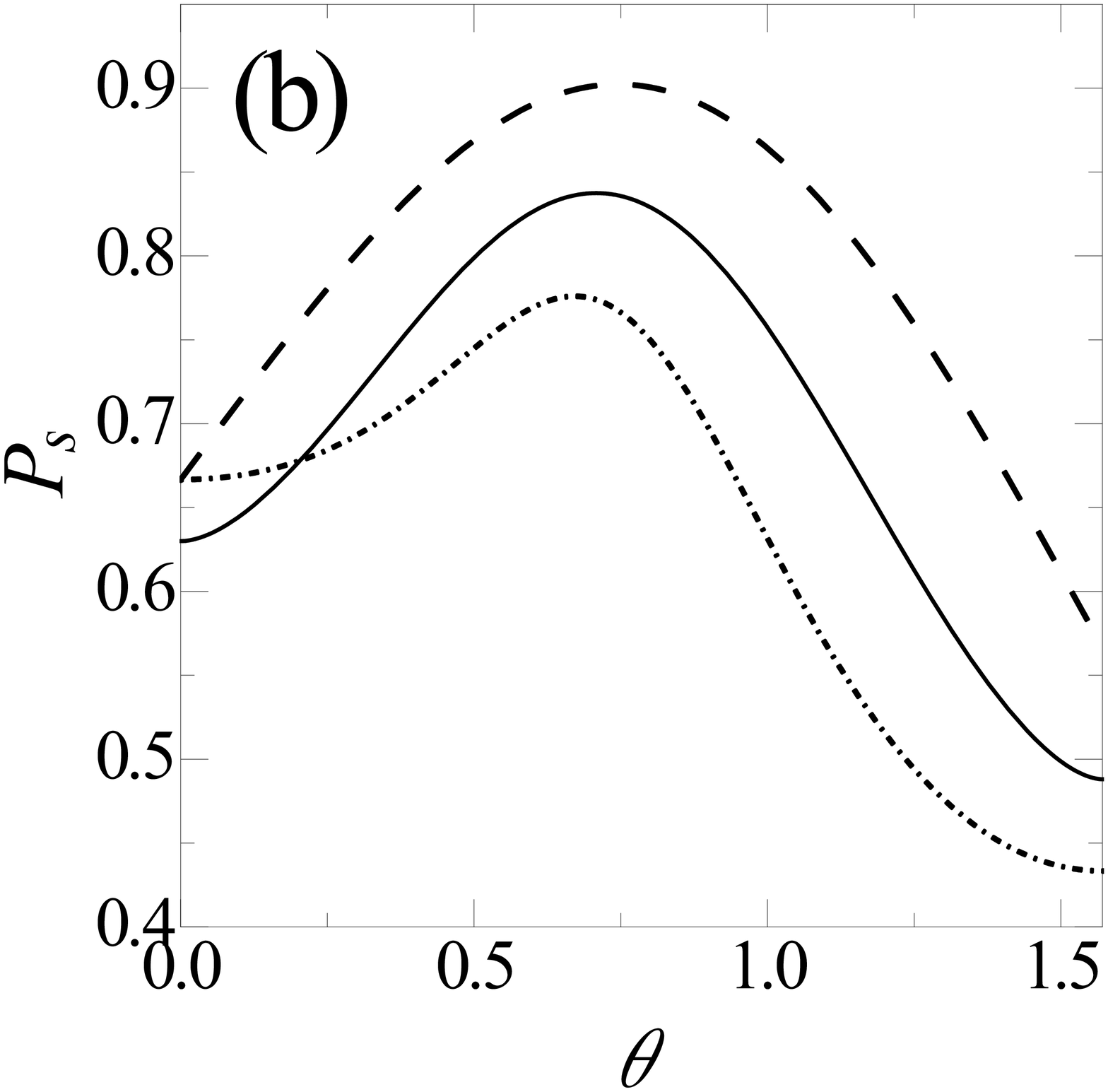}}
\caption{(a) Lower bounds for the optimal discrimination probability for  the states in \eqref{three states-1}. (b) Lower bounds  for the same set but with $\vert\psi_{2}\rangle=(\vert 0\rangle+\vert 1\rangle)/2$. The entropic lower bounds are plotted as the solid curves. For comparison, the lower bounds by the square-root measurement and the pairwise-overlap bounds are represented by the dashed and dot-dashed curves, respectively.}  \label{example}
\end{figure*}

We now consider a few exemplary sets of pure states $\{\psi_{x}\}_{x=1}^{N}$ and compare the entropic bound in \eqref{q_bound2} with the other two lower bounds.
 Let us first look at three 3-dimensional pure states with equal probabilities,
 \begin{equation} \label{three states-1}
 \begin{split}
 \vert\psi_{1}\rangle&=\sin\theta\vert0\rangle+\cos\theta\vert2\rangle,~\vert\psi_{2}\rangle=\vert1\rangle,\\
 \vert\psi_{3}\rangle&=-\sin\theta\vert0\rangle+\cos\theta\vert2\rangle.
 \end{split}
 \end{equation}
The entropic lower bound and the other two bounds are calculated and plotted in Fig.~\ref{example}(a).
In Fig.~\ref{example}(b), we present the bound values with another set of states where only
$\vert\psi_{2}\rangle$ is
replaced with  $(\vert0\rangle+\vert1\rangle)/\sqrt{2}$
from the previous case.
As shown in Fig.~\ref{example}, for both the sets of states, the square-root measurement provides the tightest bounds. The entropic bound is shown to be greater than the pairwise-overlap bound for large regions. \\

The next example of the component states is four 2-dimensional states with equal probabilities
 \begin{equation} \label{four states}
  \begin{split}
  \vert\psi_{1}\rangle&=\vert0\rangle,~ \vert\psi_{2}\rangle=\sin\theta\vert0\rangle+\cos\theta\vert1\rangle,\\
  \vert\psi_{3}\rangle&=\vert1\rangle,~ \vert\psi_{4}\rangle=\cos\theta\vert0\rangle-\sin\theta\vert1\rangle,
  \end{split}
 \end{equation}
 which satisfy $\langle\psi_{1}\vert\psi_{3}\rangle=\langle\psi_{2}\vert\psi_{4}\rangle=0$. In this case, the optimal success probability is known as ${1}/{2}$ for any $0\leqslant\theta\leqslant 2\pi$~\cite{Bae13-2}. The entropic bound gives ${1}/{2}$, and the other two lower bounds also give ${1}/{2}$.

 Finally, we consider a discrimination problem where the component states are not given, but the density operator and the number of possible states are given. Assume that a system is prepared in one of $N$ $n$-dimensional unknown states, but the system is described by the density operator $\rho=\mathbb{1}/n$. There are arbitrarily many possibilities of choosing  component states and their preparation probabilities in constructing the density operator. For instance, in the case of $N=4$ and $n=2$, we see that the four states in \eqref{four states} with $p_{1}=p_{3}={q}/{2}$ and $p_{2}=p_{4}=({1-q})/{2}$ give rise to the same density operator $\rho=\mathbb{1}/2$ for any $0\leqslant q\leqslant1$ and $0\leqslant \theta\leqslant 2\pi$. In this case, the square-root measurement cannot be specified, so one needs to take a minimization over all decompositions of the density operator to obtain a lower bound. However, the entropic bound provides $P_{s}^{*}\geqslant N^{-1}~2^{S(\mathbb{1}/n)}={1}/{2}$ regardless of the component states and the preparation probabilities.

\section{Conclusion}

We have presented an entropic lower bound for the optimal success probability of distinguishing quantum states. It provides a connection between the optimal discrimination probability  and quantum entropy, i.e.,  between a practically relevant quantity and a primary function in quantum information theory.
When the quantum states are all pure, the entropic bound is reduced to a form that requires less information for its evaluation, namely, the density operator and the number of the possible states. It shows that the von Neumann entropy can lower bound the distinguishability of $N$ pure states.


\section*{Acknowledgements}
This work was supported by the National Research Foundation of Korea(NRF) grant funded by the Korea government(MSIP) (No. 2010-0018295) and the Center for Theoretical Physics at Seoul National University.

\end{document}